\def\arxiv{}

\ifx\arxiv\undefied 

\documentclass{birkjour}

\else 

\documentclass[12pt]{article}

\usepackage[verbose,tmargin=1in,bmargin=1in,lmargin=1in,rmargin=1in]{geometry}

\fi

\usepackage[utf8]{inputenc}

\usepackage[unicode=true,pdfusetitle,
    bookmarks=true,bookmarksnumbered=false,bookmarksopen=false,
breaklinks=true,pdfborder={0 0 0},backref=false,colorlinks=true]{hyperref}
\hypersetup{linkcolor=blue,citecolor=blue,urlcolor=blue}

\usepackage{amsmath,amsthm,amsfonts}
\usepackage{enumerate}
\usepackage{slashed}
\usepackage{xcolor}

\DeclareFontFamily{U}{mathb}{\hyphenchar\font45} 
\DeclareFontShape{U}{mathb}{m}{n}{
      <5> <6> <7> <8> <9> <10> gen * mathb
      <10.95> mathb10 <12> <14.4> <17.28> <20.74> <24.88> mathb12
}{}
\DeclareSymbolFont{mathb}{U}{mathb}{m}{n}
\DeclareMathSymbol{\selfmaps}{3}{mathb}{"FD}

\usepackage{mathtools}
\usepackage{bbm}

\usepackage[normalem]{ulem}
\usepackage{cancel}
\newcommand{\red}[1]{#1}

\newcommand{\ed}{\color{black}}
\newcommand{\old}[1]{\ifx\showold\undefined\else\color{gray}\ \sout{#1}\ \ed\fi}

\def\R{{\mathbb R}}  
\def\N{{\mathbb N}}  
\def\Z{{\mathbb Z}}  
\def\C{{\mathbb C}}  

\ifx\arxiv\undefined
    
\else
    
\fi
\renewcommand{\vec}[1]{{\mathbf{#1}}}
\newcommand{\vecs}[1]{{\boldsymbol{#1}}}

\newcommand{\pol}{\mathtt{Pol}}
\newcommand{\class}{\mathtt{C}}

\newcommand{\out}{{\operatorname{out}}}
\newcommand{\inn}{{\operatorname{in}}}

\newcommand{\id}{\operatorname{id}}

\newcommand{\sk}[1]{\left\langle #1\right\rangle}

\newcommand{\range}{\operatorname{range}}

\newcommand{\supp}{\operatorname{supp}}
\newcommand{\Wedge}{\mathsf{\Lambda}}

\newcommand{\cC}{\mathcal{C}}
\newcommand{\cF}{\mathcal{F}}

\newcommand{\cH}{\mathcal{H}}

\newcommand{\cR}{\mathcal{R}}
\newcommand{\cL}{\mathcal{L}}

\newcommand{\CSigma}{{\cC_\Sigma}}

\newcommand{\HSigma}{\cH_\Sigma}

\newtheorem{thm}{Theorem}[section]
\newtheorem{cor}[thm]{Corollary}
\newtheorem{dfn}[thm]{Definition}

\begin{document}

\title{\textsc{A Perspective on External Field QED}}

\ifx\arxiv\undefined 

\author{D.-A. Deckert}
\email{deckert@math.lmu.de}
\address{%
    Mathematisches Institut der Ludwig-Maximilians-Universit\"at M\"unchen,
    Theresienstr. 39, 80333 M\"unchen, Germany
}

\author{F. Merkl}
\email{merkl@math.lmu.de}
\address{
    Mathematisches Institut der Ludwig-Maximilians-Universit\"at M\"unchen,
    Theresienstr. 39, 80333 M\"unchen, Germany
}

\else 

\author{D.-A. Deckert\thanks{deckert@math.lmu.de} \ and F.
Merkl\thanks{merkl@math.lmu.de}\\
    \small Mathematisches Institut der Ludwig-Maximilians-Universit\"at M\"unchen\\
    \small Theresienstr. 39, 80333 M\"unchen, Germany\\
}

\maketitle

\fi

\begin{abstract}
    In light of the conference \emph{Quantum Mathematical Physics} held in
    Regensburg in 2014, we give our perspective on the external field problem in
    quantum electrodynamics (QED), i.e., QED without photons in which the sole
    interaction stems from an external, time-dependent, four-vector potential.
    Among others, this model was considered by Dirac, Schwinger, Feynman, and
    Dyson as a model to describe the phenomenon of electron-positron pair
    creation in regimes in which the interaction between electrons can be
    neglected and a mean field description of the photon degrees of freedom is
    valid (e.g., static field of heavy nuclei or lasers fields).  Although it
    may appear as second easiest model to study, it already bares a severe
    divergence in its equations of motion preventing any straight-forward
    construction of the corresponding evolution operator. In informal computations
    of the vacuum polarization current this divergence leads to the need of the
    so-called \emph{charge renormalization}.  In an attempt to provide a bridge
    between physics and mathematics, this work gives a review ranging from the
    heuristic picture to our rigorous results in a way that is hopefully also
    accessible to non-experts and students.  We discuss how the evolution
    operator can be constructed, how this construction yields well-defined and
    unique transition probabilities, and how it provides a family of candidates
    for charge current operators without the need of removing ill-defined
    quantities. We conclude with an outlook of what needs to be done to identify
    the physical charge current among this family.    
\end{abstract}

\ifx\arxiv\undefined 


\maketitle

\fi

\section{Heuristic introduction}

We begin with a basic and informal introduction inspired by Dirac's original
work \cite{Dirac1934} to provide a physical intuition for the external
field QED model. Specialists among the readers are referred directly to
Section~\ref{sec:the-problem}.  As it is well-known, the free one-particle Dirac
equation, in units such that $\hbar=1$ and $c=1$,
\begin{align}
    \label{eq:free-dirac}
    (i\slashed\partial - m)\psi(x)=0,
    \qquad
    \text{for }
    \psi \in \cH = L^2(\R^3,\C^4),
\end{align}
was originally suggested to describe free motion of single electrons.  Curiously
enough, it allows for wave functions in the negative part $(-\infty,-m]$ of the
energy spectrum $\sigma(H^0)=(-\infty,-m]\cap[+m,\infty)$ of the corresponding
Hamiltonian $H^0=\gamma^0(-i\vecs{\gamma}\cdot\nabla+m)$.  
As the spectrum is not bounded from below, physicists rightfully argue
\cite{greiner_relativistic_2000} that a Dirac electron coupled to the
electromagnetic field may cascade to ever lower and lower energies by means of
radiation; the reason for this unphysical instability is that the
electromagnetic field is an open system, which may transport energy to spacial
infinity.
Other peculiarities stemming from the presence of a
negative energy spectrum are the so-called \emph{Zitterbewegung} first
observed by Schr\"odinger \cite{Schroedinger:1930} and \emph{Klein's
paradox} \cite{Klein:1929}. As Dirac demonstrated \cite{Dirac1934}, those
peculiarities can be reconciled in a coherent description when switching
from the one-particle Dirac equation \eqref{eq:free-dirac} to a many, in the
mathematical idealization even infinitely many, particle description known
as the \emph{second-quantization} of the Dirac equation.  Perhaps the most
striking consequence of this description is the phenomenon of
electron-positron pair creation, which only little later was observed
experimentally by Anderson \cite{Anderson1933}.\\

In order to get rid of peculiarities due to the negative energy states, Dirac
proposed to introduce a ``sea'' of electrons occupying all negative energy
states.  The Pauli exclusion principle then acts to prevent any additional
electron in the positive part of the spectrum to dive into the negative one. Let
us introduce the orthogonal projectors $P^+$ and $P^-$ onto the positive and
negative energy subspaces $\cH^+$ and $\cH^-$, respectively, i.e.,
$\cH^+=P^+\cH$ and $\cH^-=P^-\cH$.  \red{Dirac's heuristic picture amounts to
introducing an infinitely many-particle wave function of this sea of electrons,
usually referred to as \emph{Dirac sea},}
\begin{align}
    \label{eq:vacuum}
    \Omega = \varphi_1 \wedge \varphi_2 \wedge \varphi_3 \wedge \ldots,
    \qquad
    (\varphi_n)_{n\in\N}
    \text{ being an orthonormal basis of }
    \cH^-,
\end{align}
where $\wedge$ denotes the antisymmetric tensor product w.r.t.\  Hilbert
space $\cH$. Given a one-particle evolution operator $U:\cH\selfmaps$, such a
Dirac sea may then be evolved with an operator $\cL_U$ according to
\begin{align}
    \cL_U \Omega = U\varphi_1\wedge U\varphi_2\wedge U\varphi_3\wedge\ldots.
\end{align}
Such an ansatz may seem academic and ad-hoc. First, the Coulomb repulsion
between the electrons is neglected (not to mention radiation), second, the
choice of $\Omega$ is somewhat arbitrary.  These assumptions clearly would have
to be justified starting from a yet to be found full version of QED.  For the
time being we can only trust Dirac's intuition that the Dirac sea, when left
alone, is so homogeneously distributed that effectively every electron in it
feels the same net interaction from each solid angle, and in turn, moves freely
so that it lies near to neglect the Coulomb repulsion; see also
\cite{Deckert2010} for a more detailed discussion. Since then none of the
particle effectively ``sees'' the others, physicists refer to such a state as the
``vacuum''.  A less ad-hoc candidate for $\Omega$ would of course be the ground
state of a fully interacting theory.  Even though the net interaction
may cancel out, electrons in the ground state will be highly entangled.  The
hope in using the product state \eqref{eq:vacuum} instead, i.e., the ground
state of the free theory, to model the vacuum is that in certain regimes the
particular \red{entanglement and} motion deep down in the sea might be irrelevant.  The success of QED
in arriving at predictions which are in astonishing agreement with experimental
data substantiates this hope. \\

As a first step to introduce an interaction
one allows for an external disturbance of the 
electrons in $\Omega$ modeled by a prescribed, time-dependent,
four-vector potential $A$. This turns the one-particle Dirac equation into
\begin{align}
    \label{eq:dirac}
    (i\slashed\partial - m)\psi(x)=e\slashed A(x)\psi(x).
\end{align}
The potential $A$ may now allow for transitions of states between the subspaces
$\cH^+$ and $\cH^-$. Heuristically speaking, a state $\varphi_1\in\cH$ in the
Dirac sea $\Omega$ may be bound by the potential and over time dragged to the
positive energy subspace $\chi\in\cH^+$. For an (as we shall see,
oversimplified) example, let us assume that up to a phase the
resulting state can be represented as
\begin{align}
    \label{eq:pair}
    \Psi=
    \chi\wedge\varphi_2\wedge\varphi_3\wedge\ldots
\end{align}
in which $\varphi_1$ is missing. Due to \eqref{eq:dirac}, states in $\cH^+$ move
rather differently as compared to the ones in $\cH^-$. Thus, an electron
described by $\chi\in\cH^+$ will emerge from the ``vacuum'' and so does the
``hole'' described by the missing $\varphi_1\in\cH^-$ in the Dirac sea
\eqref{eq:pair}, which is left behind. Following Dirac, the \emph{hole} itself 
can be
interpreted as a particle, which is referred to as \emph{positron}, and both
names can be used as synonyms. If, as in this example, the electrons deeper down
in the sea are not affected too much by this disturbance, it makes sense to
switch to a more economic description. Instead of tracking all
infinitely many particles, it then suffices to describe the motion of the
electron $\chi$, of the corresponding hole $\varphi_1$, and of the net evolution
of $\Omega$ only.  Since the number of electron-hole pairs may vary over time, a
formalism for variable particle numbers is needed. This is provided by the Fock
space formalism of quantum field theory, i.e., the so-called ``second
quantization''.  One introduces a so-called \emph{creation} operator $a^*$ that
formally acts as
\begin{align}
    \label{eq:astar}
    a^*(\chi)\varphi_1\wedge \varphi_2\wedge\ldots 
    = \chi\wedge\varphi_1\wedge\varphi_2\wedge\ldots,
\end{align}
and also its corresponding adjoint $a$, which is called \emph{annihilation}
operator.  The state $\Psi$ from example in \eqref{eq:pair} can then be written
as $\Psi=a^*(\chi)a(\varphi_1)\Omega$.  With the help of $a^*$, one-particle
operators like the evolution operator $U^A$ generated by \eqref{eq:dirac} can be
lifted to an operator $\widetilde U$ on $\cF$ in a canonical way by requiring
that
\begin{align}
    \label{eq:lift}
    \widetilde U^A a^*(f)(\widetilde U^A)^{-1} = a^*(U^A f).
\end{align}
This condition determines a lift up to a phase as can be seen from the left-hand
side of \eqref{eq:lift}. Since the operator $a^*(f)$ is linear in its argument
$f\in\cH$, it is commonly split into the sum
\begin{align}
    a^*(f)=b^*(f)+c^*(f)
    \qquad
    \text{with}
    \qquad
    b^*(f):=a^*(P^+f),
    \quad
    c^*(f):=a^*(P^-f).
\end{align}
Hence, $b^*$ and $c^*$ and their adjoints are creation and annihilation
operators of electrons having positive and negative energy, respectively.  In order to be able to
disregard the infinitely many-particle wave function $\Omega$ in the notation, one
introduces the following change in language. First, the space generated by
states of the form $b^*(f_1)b^*(f_2)\ldots b^*(f_n)\Omega$ for $f_k\in\cH^+$ 
is identified with the \emph{electron 
Fock space}
\begin{align}
    \label{eq:Fe}
    \cF_e = \bigoplus_{n\in\N_0} (\cH^+)^{\wedge n}.
\end{align}
Second, the space generated by the
states of the form $c(g_1)c(g_2)\ldots c(g_n)\Omega$ for $g_k\in\cH^-$ is
identified with 
the \emph{hole Fock space}
\begin{align}
    \label{eq:Fh}
    \cF_h = \bigoplus_{n\in\N_0} (\cH^-)^{\wedge n}.
\end{align}
Note that this time the \emph{annihilation} operator of \emph{negative} energy
states is employed to generate the Fock space. To make this evident in the
notation, one usually replaces $c(g)$ by a creation operator $d^*(g)$. However,
unlike creation operators, $c(g)$ is \emph{anti-linear} in its argument
$g\in\cH^-$.  Thus, in a third step one replaces $\cH^-$ by its complex
conjugate $\overline{\cH^-}$, i.e., the set $\cH^-$ equipped with the usual
$\C$-vector space structure except for the scalar multiplication
$\cdot^\star:\C\times \overline{\cH^-}\to \overline{\cH^-}$ which is redefined
by $\lambda\cdot^\star g = \lambda^* g$ for all $\lambda\in\C$ and
$g\in\overline{\cH^-}$. This turns $\cF_h$ into
\begin{align}
    \overline{\cF_h}= \bigoplus_{n\in\N_0} (\overline{\cH^-})^{\wedge n},
\end{align}
and the hole creation operator \red{$d^*(g)=c(g)$} becomes \emph{linear} in its
argument $g\in \overline{\cH^-}$. To treat electrons and holes more
symmetrically, one also introduces the
\emph{anti-linear} charge conjugation operator $C:\cH\to\cH$,
$C\psi=i\gamma^2\psi^*$. This operator exchanges $\cH^+$ and $\cH^-$, i.e.,
$C\cH^\pm=\cH^\mp$, and thus, gives rise to a \emph{linear} map
$C:\overline{\cH^-}\to\cH^+$. A hole wave function $g\in\overline{\cH^-}$ living in
the space negative states can
then be represented by a wave function $Cg\in\cH^+$ living in the positive
energy space. \red{Our discussion of the Dirac sea above may appear to break
the charge symmetry as $\Omega$ is represented by
a sea of electrons in $\cH^-$. However, an
equivalent description that makes the charge symmetry explicit is possible by
representing the vacuum $\Omega$ through a pair of two seas, one in
$\cH^+$ and one in $\cH^-$. Nevertheless, as the charge symmetry will not play a
role in this overview we will continue using Dirac's picture with a sea of
electrons in $\cH^-$.}

By definition
\eqref{eq:astar} it can be seen that $b,b^*$ and $d,d^*$ fulfill the well-known
anti-commutator relations: 
\begin{gather}
    \begin{split}
        \{b(g),b(h)\}&= 0=\{b^*(g),b^*(h)\}, \\
        \{d(g),d(h)\}&= 0=\{d^*(g),d^*(h)\},
    \end{split}
    \qquad
    \begin{split}
        \{b^*(g),b(h)\}&=\sk{g,P^+h}\id_{\cF_e},\\
        \{d^*(g),d(h)\}&=\sk{g,P^-h}\id_{\overline{\cF_h}}.
    \end{split}
\end{gather}
The full Fock space for the electrons and positrons 
is then given by
\begin{align}
    \label{eq:fock-space}
   \cF = \cF_e \otimes \overline{\cF_h}. 
\end{align}
In this space the vacuum wave function $\Omega$ in \eqref{eq:vacuum} is
represented by $|0\rangle=1\otimes 1$ and the pair state $\Psi$ in
\eqref{eq:pair} by $a^*(\chi)d^*(\varphi_1)|0\rangle$. Thus, in this notation one 
only describes the excitations of the vacuum, i.e., those electrons that deviate
from it. The infinitely many other electrons in the Dirac sea one preferably
would like to forget about are successfully hidden in the symbol $|0\rangle$.
Here, however, the story ends abruptly. 

\subsection{The problem and a program for a cure}
\label{sec:the-problem}

For a prescribed external potential $A$, one would be inclined to compute
transition probabilities for the creation of pairs, as for example for a
transition from $\Omega$ to $\Psi$ as in \eqref{eq:vacuum} and \eqref{eq:pair},
right away.  Given the one-particle Dirac evolution operator
$U^A=U^A(t_1,t_0)$ generated by \eqref{eq:dirac} and any orthonormal basis
$(\chi_n)_n$ of $\cH^+$, the first order of perturbation of the probability of
\red{a possible pair creation} is given by
\begin{align}
    \label{eq:transition}
    \sum_{nm} \left|\sk{\chi_n,U^A\varphi_m}\right|^2
    =
    \|U^A_{+-}\|_{I_2},
\end{align}
where $I_2(\cH)$ denotes the space of bounded operators
with finite Hilbert-Schmidt norm $\|{\cdot}\|_{I_2}$, and we use the notation
$U^A_{\pm\mp}=P^\pm U^A P^\mp$.
For quite general potentials $A=(A^0,\vec A)$, it
turns out that:
\begin{thm}[\cite{Ruijsenaars1977a}]
    \label{thm:ruijsenaars}
    Term $\eqref{eq:transition} < \infty$ for all times $t_0,t_1\in\R$
    $\Leftrightarrow$
    $\vec A = 0$.
\end{thm}
In view of \eqref{eq:transition}, the transition probability is thus only defined
for external potentials $A$ that have zero spatial components $\vec A$.  
Even worse, the criterion for the well-definedness of a possible 
lift $\widetilde U$ of any
unitary one-particle operator $U$ according to \eqref{eq:lift} is given by:
\begin{thm}[\cite{Shale1965}]
    \label{thm:shale}
    There is a unitary operator $\widetilde U:\cF\selfmaps$ that
    fulfills \eqref{eq:lift}
    $\Leftrightarrow$
    $U_{+-}, U_{-+} \in I_2(\cH)$.
\end{thm}
Applying this result to the evolution operator $U^A$, \eqref{eq:transition} and
Theorem~\ref{thm:ruijsenaars} imply that the criterion in
Theorem~\ref{thm:shale} is only fulfilled for external potentials $A$ with zero
spatial components $\vec A$. Even more peculiar, the given criterion is not
gauge covariant (not to mention the Lorentz covariance).  Although the free
evolution operator $U^{A=0}$ has a lift, in the case that some spatial
derivatives of a scalar field $\Gamma$ are non-zero, the gauge transformed $U^{A=\partial\Gamma}$ does
not. This indicates that an unphysical assumption must have been made.\\

What singles out the spatial components of $A$? Mathematically, they appear in
the Hamiltonian, $H^A=\gamma^0(-i\vecs{\gamma}\cdot\vec\Delta+m)+A_0 -
\gamma^0\vecs{\gamma}\cdot \vec A$, preceded by the spinor matrix
$\gamma^0\vecs\gamma$ whereas $A_0$ is only a multiple of the identity.
Heuristically, if $\vec A$ is non-zero then the $\gamma^0\vecs\gamma$ matrix 
transforms the negative energy states $\varphi_n$ in spinor space to develop
components in $\cH^+$. There is no mechanism that would limit this
development, not even smallness of $|\vec A|$, so there is no reason why the infinite
sum \eqref{eq:transition} should be finite -- and in general this is also not
the case as Theorem~\ref{thm:ruijsenaars} shows. In other words, for $\vec A\neq0$,
instantly infinitely many electron-positron pairs are created from the vacuum
state $\Omega$. Therefore, the picture is not nearly as peaceful as suggested by
example state \eqref{eq:pair}. However, if $A$ is switched off at some later
time one can expect that almost all of these pairs disappear again, and only a
few excitations of the vacuum as in \eqref{eq:pair} will remain (hence, the name
\emph{virtual pairs} that is used by physicists).  Assuming that at initial and
final times $A=0$, it can indeed be shown that the scattering matrix $S^A$
fulfills the conditions of Theorem~\ref{thm:shale}.  
\red{The physical reason why the spatial
components are singled out is due to the use of equal-time
hyperplanes and will be discussed more geometrically in
Section~\ref{sec:polarizations}; see Theorem~\ref{thm:ident-pol-class} below.}

In conclusion, the problem lies in the fact that even the ``vacuum'' $\Omega$
consists of infinitely many particles. In the formalism of the free theory this
fact is usually hidden by the use of normal ordering. Without it the ground
state energy of $\Omega$ would be the infinite sum of all negative energies, or
the charge current operator expectation value $\sk{\Omega,\overline{a}\gamma^\mu
a \Omega}$ of the vacuum would simply be the infinite sum of all one-particle
currents $\overline\varphi_n \gamma^\mu\varphi_n$ -- both quantities that
diverge.  The rational behind the ad-hoc introduction of normal ordering of,
e.g., the charge current operator is again the assumption that in the vacuum
state these currents are effectively not observable since the net interaction
between the particles vanishes.

The incompatibility of Theorem~\ref{thm:shale} with the gauge freedom however
shows that, although the choice of $\Omega$ may be distinguished for $A=0$ by
the ground state property, it is somehow arbitrary when $A\neq 0$, and so is the
choice in the splitting of $\cH$ into $\cH^+$ and $\cH^-$, which is usually
referred to as \emph{polarization}. As a program for a cure of these divergences,
one may therefore attempt to carefully adapt the choice of the polarization
depending on the evolution of $A$ instead of keeping it fixed.  Several attempts
have been made to give a definition of a more physical polarization, one of them
being the \emph{Furry picture}. It defines the polarization according to the
positive and negative parts of the spectrum of $H^A$ given a fixed $A$.
Unfortunately, none of the proposed choices are Lorentz invariant as it is shown
in \cite{Fierz1979} since the vacuum state w.r.t.\  one of such choices in one
frame of reference may appear as a many-particle state in another. This is due
to the fact that the energy spectrum is obviously not invariant under Lorentz
boosts.

Although a fully developed QED may be able to distinguish a class of states that
can be regarded as physical vacuum states, simply by verifying the assumption
above that the net interaction between the particles vanishes, the external
field QED model has no mathematical structure to do so.  Nevertheless, whenever
a distinction between electrons and positrons by means of a polarization is not
necessary, e.g., in the case of vacuum polarization in which the exact number of
pairs is irrelevant, it should still be possible to track the time evolution
$\widetilde U^A\Omega$ and study the generated dynamics -- not only
asymptotically in scattering theory but also at intermediate times.  The choice
in admissible polarizations can then be seen to be analogous to the choice of a
convenient coordinate system to represent the Dirac seas. Since the employed
Fock space $\cF$ depends directly on the polarization of $\cH$ into $\cH^+$ and
$\cH^-$, see \eqref{eq:Fe}-\eqref{eq:Fh} and \eqref{eq:fock-space}, the standard
formalism has to be adapted to allow the Fock space to also vary according to
$A$, and the evolution operator $\widetilde U^A$ must be implemented mapping one
Fock space into another. While the idea of varying Fock space may be unfamiliar
from the non-relativistic setting, it is natural when considering a
relativistic formalism. A Lorentz boost, for example, tilts an equal-time
hyperplane to a Cauchy surface $\Sigma$ which requires a change from the
\red{standard Hilbert space $\cH=L^2(\R^3,\C^4)$} to one that is attached to
$\Sigma$, and likewise, for the corresponding Fock spaces. Hence, a Lorentz
transform will naturally be described by a map from one Fock space into another
\cite{Deckert2014}.  In the special case of equal-time hyperplanes, parts of this
program have been carried out in \cite{Langmann1996,Mickelsson1998} and
\cite{Deckert2010a}. In the former two works the time evolution operator is
\red{nevertheless} implemented on standard Fock space $\cF$ by conjugation of
the evolution operator with a convenient (non-unique) unitary
``renormalization'' transformation. In the latter work it is implemented between
time-varying Fock spaces, so-called infinite wedge spaces, and furthermore, the
degrees of freedom in the construction have been identified.  These latter
results have been extended recently to allow for general Cauchy surfaces in
\cite{Deckert2014,Deckert2015} and are presented in
Section~\ref{sec:polarizations}. All these results ensure the existence of an
evolution operator by a quite abstract argument. Therefore, we review a
construction of it in Section~\ref{sec:wedge-spaces} based on
\cite{Deckert2010a}.  It utilizes a notation that is very close to Dirac's
original view of a sea of electrons as in \eqref{eq:vacuum}. Though it is canonically
equivalent to the Fock space formalism, it provided us a more intuitive view of
the problem and helped in identifying the degrees of freedom involved in the
construction.  In Section~\ref{sec:phase} we conclude with a discussion of the
unidentified phase of the evolution operator and its meaning for the charge
current in. Beside the publications cited so far, there are several recent
contributions which also take up on  Dirac's original idea.  As a more
fundamental approach we want to mention the one of the so-called ``Theory of
Causal Fermion Systems'' \cite{FinsterBook,Finster2015a,Finster2015b}, which is based on a reformulation of quantum
electrodynamics from first principles. The phenomenon of adiabatic pair creation
was treated rigorously in \cite{Pickl2008}. Furthermore, there is a series of
works treating the Dirac sea in the Hartree-Fock approximation. The most general
is \cite{Gravejat2013} in which the effect of vacuum polarization was
treated self-consistently for static external sources.

\section{Varying Fock spaces}
\label{sec:polarizations}

In order to better understand why the spatial components of $A$ had been singled
out in the discussion above, it is helpful to consider the Dirac evolution not
only on equal-time hyperplanes but on more general Cauchy surfaces.
\begin{dfn}\label{def:cauchy-surface}
    A Cauchy surface $\Sigma$ in $\R^4$ is a smooth, 3-dimensional
    submanifold of $\R^4$ that fulfills the following two conditions:
    \begin{enumerate}
        \item Every inextensible,  two-sided, time- or light-like, continuous path
            in $\R^4$ intersects $\Sigma$ in a unique point.
        \item For every $x\in\Sigma$, the tangent space $T_x\Sigma$ of
            $\Sigma$ at $x$ is
            space-like.
    \end{enumerate} 
\end{dfn}
To each Cauchy surface $\Sigma$ we associated a Hilbert space $\HSigma$.
\begin{dfn}
    \label{def:HSigma}
    Let $\HSigma=L^2(\Sigma,\C^4)$ denote the vector space of all
    4-spinor valued measurable functions $\phi:\Sigma\to \C^4$ (modulo changes
    on null sets) having a finite norm $\|\phi\|=\sqrt{\sk{\phi,\phi}}<\infty$
    w.r.t.\ the scalar product 
    \begin{equation}
        \label{eq:scalar-product}
        \sk{\phi,\psi}
        =\int_{\Sigma}\overline{\phi(x)}i_\gamma(d^4x)\psi(x).
    \end{equation}
    Here, $i_\gamma(d^4x)$ denotes the contraction of the volume form
    $d^4x=dx^0\wedge dx^1\wedge dx^2\wedge dx^3$ with the spinor-matrix valued
    vector $\gamma^\mu$, $\mu=0,1,2,3$. 
    The corresponding dense subset of smooth and compactly supported functions
    will be denoted by $\CSigma$.
\end{dfn}
The well-posedness of the initial value problem related to \eqref{eq:dirac} for
initial data on Cauchy surfaces has been studied in the literature;
e.g., see \cite{john:82,taylor:11} for general hyperbolic systems and more
specifically for wave equations on Lorentzian manifolds \cite{dimock:82},
\cite{baer:2007}, \cite{ringstroem:09}, \cite{Finster:2012}, and
\cite{Derezinski:13}. For the purpose of our study we furthermore introduced
generalized Fourier transforms for the Dirac equation in \cite{Deckert2014} and
extended the standard Sobolev and Paley-Wiener methods in $\R^n$ to the geometry
given by the Cauchy surfaces and the mass shell of the Dirac equation. These
methods were required for the analysis of solutions. They play along nicely 
with Lorentz and gauge transforms and allow for the introduction of an
interaction picture. As a byproduct, these methods also ensure existence, uniqueness, and
causal structure of strong solutions.   Since we avoid technicalities in this
paper, we assume $A$ is a smooth and compactly supported (although sufficient
strong decay would be sufficient), and the following theorem will suffice to
discuss the one-particle Dirac evolution.
\begin{thm}[Theorem~2.23 in \cite{Deckert2014}] 
    \label{thm:one-part-U}
    Let $\Sigma,\Sigma'$ be two Cauchy surfaces and $\psi_\Sigma\in \CSigma$ the
    initial data.  There is a unique strong solution
    $\psi\in\cC^\infty(\R^4,\C^4)$ to \eqref{eq:dirac} being 
    supported in the forward and backward
    light cone of $\supp \psi_\Sigma$ such that
    $\psi|_{\Sigma}=\psi_\Sigma$ holds.  Furthermore, there is an isometric
    isomorphism
    $U^A_{\Sigma'\Sigma}:\CSigma\to\cC_{\Sigma'}$ fulfilling
    $\psi|_{\Sigma'}=U^A_{\Sigma'\Sigma}\psi_\Sigma$. Its unique extension to a
    unitary map $U^A_{\Sigma'\Sigma}:\HSigma\to\cH_{\Sigma'}$ is denoted by the same symbol.
\end{thm}
Similarly to the standard Fock space
\eqref{eq:fock-space} we define the Fock space
for a Cauchy surface on the basis of a polarization.
\begin{dfn}
    Let $\pol(\HSigma)$ denote the set of all closed, linear subspaces
    $V\subset \HSigma$ such that $V$ and $V^\perp$ are both infinite
    dimensional. Any $V\in \pol(\HSigma)$ is called a {\em polarization} of
    $\HSigma$.  For $V\in \pol(\HSigma)$, let $P_\Sigma^V:\HSigma\to V$
    denote the orthogonal projection of $\HSigma$ onto $V$.
\end{dfn}
The Fock space attached to Cauchy surface $\Sigma$
and corresponding to polarization $V\in\pol(\HSigma)$ is defined by
\begin{align}
    \label{eq:Fock-space}
    \cF(V,\Sigma) 
    :=
    \bigoplus_{c\in\Z} \cF_c(V,\HSigma),
    \qquad
    \cF_c(V,\Sigma) 
    :=
    \bigoplus_{\substack{n,m\in\N_0\\c=m-n}} (V^\perp)^{\wedge n} \otimes
    \overline V^{\wedge m}.
\end{align}
Note that the standard Fock space is included in this definition by choosing
$\Sigma=\{0\}\times\R^3$ and $V=\cH^-$.

Given two Cauchy surfaces $\Sigma$ and $\Sigma'$, polarizations $V\in\pol(\HSigma)$
and $V'\in\pol(\cH_{\Sigma'})$, and the one-particle evolution operator
$U^A_{\Sigma'\Sigma}:\HSigma\to\cH_{\Sigma'}$, we need a condition analogous to
\eqref{eq:lift} that allows us to find an evolution operator $\widetilde
U^A_{V',\Sigma';V,\Sigma}:\cF(V,\Sigma)\to\cF(V',\Sigma')$.  For the discussion,
let $a^*_\Sigma$ and $a_\Sigma$ denote the corresponding creation and
annihilation operators on any $\cF(W,\Sigma)$ for $W\in\pol(\HSigma)$; 
note that the defining expression of $a^*$ in \eqref{eq:astar}
does  not depend on the choice of a polarization $W$.
In this notation, the
lift requirement reads
\begin{align} 
    \widetilde U^A_{V',\Sigma';V,\Sigma} 
    \; a^*_{\Sigma}(f) \; 
    \left(\widetilde U^{A}_{V',\Sigma';V,\Sigma}\right)^{-1} =
    a^*_{\Sigma'}(U^A_{\Sigma'\Sigma}f),
    \qquad
    \forall\,f\in\HSigma.
    \label{eq:lift-condition}
\end{align}
The condition under which such a lift of the one-particle evolution operator
$U^A_{\Sigma'\Sigma}$ exists can be inferred
from a slightly rewritten version of the Shale-Stinespring Theorem~\ref{thm:shale}:
\begin{cor}
    \label{cor:gen-shale}
    Let $\Sigma,\Sigma'$ be Cauchy surfaces, $V\in\pol(\HSigma)$, and
    $V'\in\pol(\cH_{\Sigma'})$. Then the following statements are equivalent:
    \begin{enumerate}
        \item There is a unitary operator $\widetilde
            U^A_{V'\Sigma';V,\Sigma}:\cF(V,\Sigma)\to\cF(V',\Sigma')$ which fulfills
            (\ref{eq:lift-condition}).
        \item The off-diagonals $P^{ {V'}^\perp}_{\Sigma'}
            U^A_{\Sigma'\Sigma} P^V_\Sigma$ and
            $P^{V'}_{\Sigma'} U^A_{\Sigma'\Sigma}
            P^{V^\perp}_\Sigma$ are Hilbert-Schmidt operators.
    \end{enumerate}
\end{cor}
Note again that if such a lift exists, its phase is not fixed by
\eqref{eq:lift-condition} and the corollary above does not provide any information
about it. Therefore, we will discuss a direct construction of the lifted operator
$\widetilde U^A_{V'\Sigma';V,\Sigma}$ in Section~\ref{sec:wedge-spaces}, which
makes the involved degrees of freedom apparent.

Coming back to the question
which polarizations $V\in\pol(\HSigma)$ and $V'\in\pol(\cH_{\Sigma'})$
guarantee the existence of a lifted evolution operator $\widetilde
U^A_{\Sigma'\Sigma}:\cF(V,\Sigma)\to\cF(V',{\Sigma'})$, one readily finds
a trivial choice. Let us pick a Cauchy surface $\Sigma_\inn$ in the
remote past fulfilling:
\begin{align}
    \Sigma_\inn \text{ is a Cauchy surface such that }
    \supp A\cap\Sigma_\inn=\emptyset.
\end{align}
When transporting the standard polarization along with the Dirac evolution 
we get 
\begin{align}
    \label{eq:interpolation}
    V=U^A_{\Sigma\Sigma_\inn} \, 
    P^-_{\Sigma_\inn} \cH_{\Sigma_\inn} \in \pol(\HSigma),
    \qquad
    V'=U^A_{\Sigma'\Sigma_\inn}\,
    P^-_{\Sigma_\inn}\cH_{\Sigma_\inn}\in\pol(\cH_{\Sigma'}),
\end{align}
which automatically fulfills condition (b) of
Theorem~\ref{cor:gen-shale} as then the off-diagonals
\red{$(U^A_{\Sigma\Sigma_\inn})_{\pm\mp}$} become zero. This choice
is usually called the \emph{interpolation picture}.  Its drawback 
is that the polarizations $V$ and $V'$ depend on the whole history of $A$ between
$\Sigma_\inn$ and $\Sigma$ and $\Sigma'$. Moreover, such $V$ and $V'$ are rather
implicit. Luckily, there are other choices. 
Statement (b) in Theorem~\ref{cor:gen-shale} allows to differ from
the projectors $P^V_\Sigma$ and $P^{V'}_{\Sigma'}$ by a Hilbert-Schmidt operator.
Hence, all admissible polarizations can be collected and characterized by means
of the
following classes:
\begin{dfn}
    \label{def:pol-classes}
    For a Cauchy surface $\Sigma$ we define the class
    \begin{align}
        \class_\Sigma(A)
        :=
        \left\{ W \in \pol(\HSigma) \,\big|\, 
            W\approx U_{\Sigma\Sigma_\inn}^A \cH_{\Sigma_\inn}^- \right\}
    \end{align}
    where for $V, W\in \pol(\HSigma)$, $V\approx W$ means that the difference of
    the corresponding orthogonal projectors
    $P_\Sigma^V-P_\Sigma^{W}$ is a Hilbert-Schmidt operator.
\end{dfn}
As simple implication of
Corollary~\ref{cor:gen-shale} one gets:
\begin{cor}
    Let $\Sigma,\Sigma'$ be Cauchy surfaces and polarizations $V\in C_\Sigma(A)$
    and $W\in C_{\Sigma'}(A)$. Then up to a phase there is a unitary operator
    $\widetilde U^A_{\Sigma'\Sigma}:\cF(V,\HSigma)\to\cF(W,\cH_{\Sigma'})$
    obeying \eqref{eq:lift-condition}.
\end{cor}
We emphasize again that any other possible polarization than the choice in
\eqref{eq:interpolation} is comprised in the respective class $C_\Sigma(A)$ as
Corollary~\ref{cor:gen-shale} only allows for the freedom encoded in the
equivalence relation ${\approx}$. Although the polarization
\eqref{eq:interpolation} depends on the history of the evolution it turns out
that the classes $C_\Sigma(A)$ are independent thereof. The sole dependence of
the classes $C_\Sigma(A)$ is on the tangential components of $A$, which can be
stated as follows.
\begin{thm}[Theorem~1.5 in \cite{Deckert2015}]
    \label{thm:ident-pol-class}
   Let $\Sigma$ be a Cauchy surface and let $A$ and $\widetilde A$ be two smooth
   and compactly supported
   external fields. Then
   \begin{align}
       C_\Sigma(A)=C_\Sigma(\widetilde A)
       \qquad
       \Leftrightarrow
       \qquad
       A|_{T\Sigma}=\widetilde A|_{T\Sigma},
   \end{align}
   where $A|_{T\Sigma}=\widetilde A|_{T\Sigma}$ means that for all $x$ in
   $\Sigma$ and all vectors
   $y$ in the tangent space $T_x\Sigma$ of $\Sigma$ at $x$, the relation
   $A_\mu(x)y^\mu = \widetilde A_\mu(x)y^\mu$ holds.
\end{thm}
This theorem is a generalization of Ruijsenaar's result \cite{Ruijsenaars1977}
and helps to understand why on equal-time hyperplanes the spatial components
of $A$ appeared to play such a special role. The spatial components $\vec A$ are the
tangential ones w.r.t. such Cauchy surfaces.
Furthermore, the classes $C_\Sigma(A)$
transform nicely under Lorentz and gauge transformations:
\begin{thm}[Theorem~1.6 in \cite{Deckert2015}]
    \label{thm:lorentz-gauge}
    \mbox{}
    \begin{enumerate}[(i)]
        \item Consider a {Lorentz transformation}
            given by $L^{(S,\Lambda)}_\Sigma:\HSigma\to\cH_{\Lambda\Sigma}$
            for a spinor transformation matrix $S\in \C^{4\times 4}$
            and an associated proper orthochronous Lorentz transformation
            matrix $\Lambda\in \operatorname{SO}^\uparrow(1,3)$, see for
            example \cite[Section~2.3]{Deckert2014}.
            Then:
            \begin{align}
                V\in\class_\Sigma(A)
                \qquad
                \Leftrightarrow
                \qquad
                L^{(S,\Lambda)}_\Sigma V\in
                \class_{\Lambda\Sigma}(\Lambda A(\Lambda^{-1}\cdot)).
            \end{align}
        \item 
            Consider a {gauge transformation} 
            $A'=A+\partial \Gamma$ for some 
            $\Gamma\in\cC^\infty_c(\R^4,\R)$
            given by the multiplication operator
            $e^{-i\Gamma}:\HSigma\to\HSigma$,
            $\psi\mapsto \psi'=e^{-i\Gamma}\psi$.
            Then:
            \begin{align}
                V\in\class_\Sigma(A)
                \qquad
                \Leftrightarrow
                \qquad
                e^{-i\Gamma}V\in
                \class_\Sigma(A+\partial\Gamma).
            \end{align}
    \end{enumerate}
\end{thm}
As an analogy from geometry one could think of the particular polarization as a
particular choice of coordinates to represent the Dirac sea.
Corollary~\ref{cor:gen-shale} and Theorem~\ref{thm:lorentz-gauge} explain why
gauge transformations that introduce spatial components in the external fields
do not comply with the condition to the Shale-Stinespring
Theorem~\ref{thm:shale} in which the \red{``coordinates'' $\cH^+$ and $\cH^-$
were fixed.}

The key idea in the proofs of Theorem~\ref{thm:ident-pol-class} and
\ref{thm:lorentz-gauge} is to guess a simple enough operator
$P^A_\Sigma:\HSigma\selfmaps$ depending only
on the restriction $A|_\Sigma$ so
that
\begin{align}
    \label{eq:key-prop}
    U^A_{\Sigma\Sigma_\inn}P^-_{\Sigma_\inn} U^A_{\Sigma_\inn\Sigma}
    -
    P^A_\Sigma \in I_2(\HSigma),
    \qquad
    \text{and}
    \qquad
    (P^A_\Sigma)^2 - P^A_\Sigma \in I_2(\HSigma).
\end{align}
The claims about the properties of the polarization classes
$C_\Sigma(A)$ can then be inferred directly from the properties of $P^A_\Sigma$.
This is due to the fact that \eqref{eq:key-prop} is compatible with the
Hilbert-Schmidt operator freedom encoded in the ${\approx}$ equivalence
relation.
The intuition behind the guess of $P^A_\Sigma$ used in the proofs presented in
\cite{Deckert2015} comes from the gauge transform.
Imagine the special situation in which an external potential $A$ could be gauged
to zero, i.e., $A=\partial\Gamma$ for a given scalar field $\Gamma$. In this
case $e^{-i\Gamma} P^-_{\Sigma} e^{i\Gamma}$ is a good candidate for
$P^A_\Sigma$.
Now in the case of general external potentials $A$ that cannot be attained by a
gauge transformation of the zero potential, the idea is to implement gauge
transforms locally at each space-time point. For example, if $p_-(x,y)$ denotes
the informal integral kernel of the operator $P^-_\Sigma$, one could try to define
$P^A_\Sigma$ as the operator corresponding to the informal kernel
$p^A(x,y)=e^{-i\lambda_A(x,y)}p_-(x,y)$ for the choice
$\lambda_A(x)=A(x)_\mu(y-x)^\mu$. The effect of $\lambda_A(x,y)$ on the
projector can be interpreted as a local gauge transform of $p_-(x,y)$ from the
zero potential to the potential $A_\mu(x)$ at space-time point $x$.  A careful
analysis of $P^A_\Sigma$, which was conducted in Section~2 of \cite{Deckert2015},
shows that $P^A_\Sigma$ fulfills \eqref{eq:key-prop}.

Finally, given Cauchy surface $\Sigma$, there is also an explicit representative
of the polarization class $C_\Sigma(A)$ which can be given in
terms of the bounded operator $Q^A_\Sigma:\HSigma\selfmaps$ defined by
\begin{align}\label{eq:V_Sigma}
    Q_\Sigma^A
    :=
    P^+_{\Sigma} (P^A_\Sigma - P^-_\Sigma) P^-_{\Sigma}
    -
    P^-_{\Sigma} (P^A_\Sigma - P^-_\Sigma) P^+_{\Sigma}.
\end{align}
With it, the polarization class can be identified as follows:
\begin{thm}[Theorem~1.7 in \cite{Deckert2015}]
    \label{thm:representative}
    Given Cauchy surface $\Sigma$,
    $\class_\Sigma(A)=\left[e^{Q_\Sigma(A)}
    \cH_{\Sigma}^-\right]_{\approx}$.
\end{thm}

The implications of these results on the physical picture can be seen as
follows. The Dirac sea on Cauchy surface $\Sigma$ can be described in any Fock
space $\cF(V,\HSigma)$ for any choice of polarization $V\in C_\Sigma(A)$. The
polarization class $C_\Sigma(A)$ is uniquely determined by the tangential
components of the external potential $A$ on $\Sigma$.  When regarding the Dirac
evolution from one Cauchy surface $\Sigma$ to $\Sigma'$, another choice of
``coordinates'' $V'\in C_{\Sigma'}(A)$ has to be made. Then one yields an
evolution operator $\widetilde
U_{\Sigma'\Sigma}^A:\cF(V,\HSigma)\to\cF(V',\cH_{\Sigma'})$ which is unique up
to an arbitrary phase. Transition probabilities $|\langle \Psi,\widetilde
U^A_{\Sigma'\Sigma}\Phi\rangle|^2$ for $\Psi\in\cF(V',\cH_{\Sigma'})$ and
$\Phi\in\cF(V,\HSigma)$ are well-defined and unique without the need of a
renormalization method.  Finally, for a family of Cauchy surfaces
$(\Sigma_t)_{t\in\R}$ that interpolates smoothly between $\Sigma$ and $\Sigma'$
one can also infer an infinitesimal version of how the external potential $A$
changes the polarization in terms of the flow parameter $t$; see Theorem~2.6 in
\cite{Deckert2015}.

We remark that the kernel of the orthogonal projector corresponding to a
polarization in $\class_\Sigma(A)$, which can be interpreted as a
distribution, is frequently called \emph{two-point function}. Two
kernels belonging to two polarizations in the same class $\class_\Sigma(A)$ may
differ by a square-integrable kernel. This stands in contrast to the so-called
Hadamard property (see, e.g., \cite{kay_theorems_1991}) which allows changes
with $\cC^\infty$ kernels as freedom in two-point functions.

\section{An explicit construction of the evolution operator}
\label{sec:wedge-spaces}

The argument in Section~\ref{sec:polarizations} that ensures the existence of
dynamics on varying Fock spaces is quite abstract. In this section we present a
more direct approach that is also closer to Dirac's original picture in
describing infinite particle wave functions like in \eqref{eq:vacuum}. As
discussed, the infinitely many particles are also present in the usual Fock
space formalism but commonly hidden by use of normal ordering. But since the
very obstacle in a straight-forward construction of the evolution operator is
due to their presence, it seems to make sense to work with a formalism that
makes them apparent. One such formalism, introduced in Section~2 of
\cite{Deckert2010a}, employs so-called infinite wedge spaces and will be used in
the following.\\

To leave our discussion general, let $\cH$ be a one-particle Hilbert
space (e.g., $\cH=\cH_{\Sigma}$ as in Section~\ref{sec:polarizations}) and let
$V\in\pol(\cH)$ be a polarization thereof. The Dirac sea corresponding to that
choice of polarization can be represented, using any orthonormal basis
$(\varphi_n)_{n\in\N}$ that spans $V$, by the infinite wedge product
\begin{align}
    \label{eq:some-sea}
    \Wedge\Phi=\varphi_1\wedge\varphi_2\wedge\varphi_3\wedge\ldots,
\end{align}
i.e., the anti-symmetric product of all wave functions $\varphi_n$, $n\in\N$.  Slightly
more general, it suffices if $(\varphi_n)_{n\in\N}$ is only \emph{asymptotically}
orthonormal in the sense that the infinite matrix
$(\sk{\varphi_n,\varphi_m})_{n,m\in\N}$ has a (Fredholm) determinant, i.e., that
it differs from the identity only by a
matrix that has a trace. The reason for this property will become clear when introducing
the scalar product of two infinite wedge products.

In order to keep the formalism short, we encode the basis $(\varphi_n)_{n\in\N}$
by a bounded linear operator 
\begin{align}
    \Phi:\ell\to\cH,
    \qquad
    \Phi\, e_n = \varphi_n
\end{align}
on a Hilbert space $\ell$.
The role of $\ell$ is only that of an index space, and one example we
have in mind is $\ell=\ell^2(\N)$, i.e., the space of square summable sequences
where the vectors $e_n$, $n\in\N$, denote the canonical basis. In this language,
the asymptotic orthonormality requirement from above can be rewritten as
$\Phi^*\Phi\in \id_\ell+I_1(\ell)$, where $I_1(\ell)$ is the space of bounded
linear maps $\ell\to\ell$ which have a trace, the so-called \emph{trace class}. We
will also write $\Wedge \Phi = \varphi_1\wedge\varphi_2\wedge\ldots$ which
denotes the infinite wedge product \eqref{eq:some-sea} and refer to all such
$\Phi$ as \emph{Dirac seas}.

Given another Dirac sea $\Psi$ with $\psi_n=\Psi e_n$, $n\in\N$,
the pairing that will later become a scalar product
\begin{align}
    \label{eq:pairing}
    \sk{\Wedge \Psi,\Wedge \Phi}
    =
    \sk{\psi_1\wedge \psi_2\wedge\ldots,\varphi_1\wedge
    \varphi_2\wedge\ldots}
    =
    \det (\sk{\psi_n,\varphi_m})_{nm}
    =
    \det \Psi^*\Phi
\end{align}
is well-defined if $\Psi^*\Phi$ has a determinant, which is the case if
$\Psi^*\Phi\in \id_\ell+I_1(\ell)$.  Thus, it makes sense to build a Fock space,
referred to as ``infinite wedge space $\cF_{\Wedge \Phi}$'', based on a basis encoded by
$\Phi$.  It is defined by the completion w.r.t.\ the pairing \eqref{eq:pairing}
of the space of \emph{formal} linear combinations of all such $\Psi$; see
Section~2.1 in \cite{Deckert2010a} for a rigorous construction. This space
consists of the sea wave function $\Wedge \Phi$, its excitations $\Wedge\Psi$
that form a generating set, and superpositions thereof. An example excitation
analogous to \eqref{eq:pair} representing an electron-positron pair with
electron wave function $\chi\in V^\perp$ and positron wave function
$\varphi_1\in V$
is given by 
\begin{align}
    \Wedge \Psi = \chi\wedge\varphi_2\wedge\varphi_3\wedge \varphi_4\wedge\ldots.
\end{align}
Note, however, that mathematically $\Phi$ is not distinguished as ``the
one vacuum'' state as it turns out that $\cF_{\Wedge\Phi}=\cF_{\Wedge\Psi}$ if
and only if $\Psi^*\Phi$ has a determinant, i.e., if the scalar product
$\sk{\Wedge \Psi,\Wedge\Phi}$ in \eqref{eq:pairing} is well-defined. This is due
to the that fact $\Psi\sim\Phi:\Leftrightarrow\Psi^*\Phi\in \id_\ell+I_1(\ell)$
is an equivalence relation on the set of all Dirac seas; see Corollary~2.9 in
\cite{Deckert2010a}.

Next, let us consider another one-particle Hilbert space $\cH'$ und a
one-particle unitary operator $U:\cH\to\cH'$ such as the one-particle Dirac
evolution operator $U^A_{\Sigma'\Sigma}$.  To infer from this a corresponding
evolution of the Dirac seas, we define a canonical operation from the left  as
follows
\begin{align}
    \label{eq:left-op}
    \cL_U:\cF_{\Wedge\Phi} \to \cF_{\Wedge U\Phi},
    \qquad
    \cL_U \, \Wedge \Psi := \Wedge U\Psi =
    (U\psi_1)\wedge(U\psi_2)\wedge\ldots.
\end{align}
Here, $\Psi$ is taken from the generating set of Dirac seas fulfilling
$\Psi^*\Phi\in 1+I_1(\ell)$; 
see
Section~2.2 in \cite{Deckert2010a}. That the range of $\cL_U$ is $\cF_{\Wedge U\Phi}$
is due to the fact that $\Psi^*\Phi$ has a determinant if and only if
$(U\Psi)^*(U\Phi)$ does. Such a map
$\cL_U$ represents an evolution operator from one infinite wedge space into
another that in the sense of \eqref{eq:astar} also complies with the previously
discussed lift condition \eqref{eq:lift}.  

Nevertheless, the construction of the evolution operator for the Dirac seas does
not end here because the target space $\cF_{\Wedge U\Phi}$ in \eqref{eq:left-op}
is completely implicit, and hence, $\cL_U$ alone is not very helpful.  On the
contrary, relying on the observations made in
Section~\ref{sec:polarizations}, physics should allow us to decide beforehand
between which infinite wedge spaces the evolution operator should be implemented.
Consider the example situation of
\begin{align}
    \label{eq:example}
    \begin{gathered}
        \text{an evolution operator } U=U^A_{\Sigma'\Sigma} \text{ from
            Theorem~\ref{thm:one-part-U}},
        \\
        \cH=\HSigma,
        \quad
        V\in\pol(\HSigma),
        \quad
        \Phi:\ell\to\cH_{\Sigma} \text{ such that } \range\Phi=V,
        \\
        \cH'=\cH_{\Sigma'},
        \quad
        V'\in\pol(\cH_{\Sigma'}),
        \quad
        \Phi':\ell'\to\cH_{\Sigma'} \text{ such that } \range\Phi'=V'.
    \end{gathered}
\end{align}
In this situation one would wish for an evolution operator of the form
$\widetilde U: \cF_{\Wedge \phi}\to\cF_{\Wedge\Phi'}$ instead of $\widetilde U:
\cF_{\Wedge \phi}\to\cF_{\Wedge U\Phi}$.  If we are not in the lucky case
$\cF_{\Phi'}=\cF_{\Wedge U\Phi}$, there are two ways in which the equality may
fail. First, Corollary~\ref{cor:gen-shale} suggests that polarization $V$ and
$V'$ must be elements of the appropriate polarization classes, more precisely,
$V\in C_\Sigma(A)$ and $V'\in C_{\Sigma'}(A)$.  However, there is a more subtle
obstacle as for $\cF_{\Phi'}=\cF_{\Wedge U\Phi}$ to hold we need to ensure that
$\sk{\Phi',U\Phi}$ is well-defined, which even for $\ell=\ell'$ and admissible
$V$ and $V'$ does need not to be the case. Thus, in general $U\Phi$ and $\Phi'$ belong
to entirely different infinite wedge spaces as the
choice of orthonormal bases encoded in $\Phi$ and $\Phi'$ was somehow
arbitrary. However, let $\Psi:\ell\to\cH'$ be another Dirac sea with
$\range \Psi=V'$, then there is a unitary $R:\ell'\to\ell$ such that $\Phi'=\Psi
R$. The action of $R$ gives rise to a unitary operation from the right $\cR_R$
characterized
by
\begin{align}
    \cR_R:\cF_{\Wedge\Psi}\to\cF_{\Wedge\Psi R},
    \qquad
    \cR_R \, \Wedge \tilde \Psi = \Wedge ( \tilde\Psi R )
\end{align} 
for all $\tilde\Psi:\ell\to\cH'$ in the generating system of $\cF_{\Wedge\Psi}$,
which connects the infinite wedge spaces $\cF_{\Wedge\Psi}$ and
$\cF_{\Wedge\Phi'}$.
The spaces $\cF_{\Wedge\Psi}$ and $\cF_{\Wedge \Phi'}$ coincide if and
only if \red{$\ell=\ell'$ and} $R$ has a determinant.  Slightly more generally,
it suffices if $R$ is
only \emph{asymptotically} unitary in the sense that $R^*R$ has a non-zero
determinant. Then the operation from the right $\det(R^*R)^{-1/2}\cR_R$ is
unitary. Whether there is a unitary $R:\ell'\to\ell$ in the situation of example
\eqref{eq:example} above such that $\cF_{\Wedge U\Phi R}=\cF_{\Wedge \Phi'}$ is
answered by the next theorem. It can be seen as yet another version of
the Shale and Stinespring's Theorem:
\begin{thm}[Theorem~2.26 of \cite{Deckert2010a}]
    \label{thm:our-shale}
    Let $\cH,\ell,\cH',\ell'$ be Hilbert spaces, $V\in\pol(\cH)$ and
    $V'\in\pol(\cH')$ polarizations, \red{$\Phi:\ell\to\cH$ and
    $\Phi':\ell'\to\cH'$} Dirac seas such that
    $\range \Phi=V$ and $\range \Phi'=V'$. Then the following statements are
    equivalent:
    \begin{enumerate}
        \item The off-diagonals $P^{{V'}^\perp}UP^V$ and
            $P^{{V'}}UP^{V^\perp}$ are Hilbert-Schmidt operators.
        \item There is a unitary $R:\ell'\to\ell$ such that
            $\cF_{\Wedge\Phi'}=\cF_{\Wedge U\Phi R}$.
    \end{enumerate}
\end{thm}
Coming back to the example \eqref{eq:example} from above, in the case $V\in
C_{\Sigma}(A)$ and $V'\in C_{\Sigma'}(A)$, i.e., that the chosen polarization belong
to the admissible classes of polarizations, condition (a) of
Theorem~\ref{thm:our-shale} is fulfilled, which implies the existence of a
unitary map $R:V'\to V$ such that the evolution operator
\begin{align}
    \label{eq:evolution}
    \widetilde U^A_{V,\Sigma;V'\Sigma'}:
    \cF_{\Wedge\Phi}
    \to
    \cF_{\Wedge\Phi'},
    \qquad
    \widetilde U^A_{V,\Sigma;V'\Sigma'}
    =
    \cR_R \circ \cL_{U^A_{\Sigma'\Sigma}}
\end{align}
is well-defined and unitary. An immediate question is of course how many such
maps exist, and it turns out that any other operation from the right $\cR_{R'}$
for which $\cR_{R'}\circ\cL_U:\cF_{\Wedge\Phi}\to\cF_{\Wedge\Phi'}$ is
well-defined and unitary fulfills ${\widetilde
U}^A_{V,\Sigma;V'\Sigma'}=e^{i\theta}\,\cR_{R'}\circ\cL_U$ for some
$\theta\in\R$; see \cite[Corollary~2.28]{Deckert2010a}.  Now $\Phi$ and $\Phi'$
are Dirac seas in which all states in $V$ and $V'$ are
occupied, respectively.  A canonical choice for their representation is
to choose $\ell=V$,
$\ell'=V'$, and to define the inclusion maps
$\Phi:V\xhookrightarrow{}\cH_{\Sigma}$, $\Phi v = v$ for all $v\in V$, and
$\Phi':V'\xhookrightarrow{}\cH_{\Sigma'}$, $\Phi'v'=v'$ for all $v'\in V'$. In
this case there is a canonical isomorphism between the spaces $\cF_{\Wedge\Phi}$
and $\cF_{V,\Sigma}$ as well as between $\cF_{\Wedge\Phi'}$ and
$\cF_{V',\Sigma'}$. Hence, we are again in the situation of
Corollary~\ref{cor:gen-shale}. We can identify the
evolution of the Dirac seas only up to a
phase $\theta\in\R$. However, now we have a more direct construction at hand
which identifies the involved degrees of freedom:
\begin{enumerate}
    \item The choice of particular polarizations $V\in C_\Sigma(A)$ and $V'\in
        C_{\Sigma'}(A)$.
    \item The choice of particular bases encoded in $\Phi$ and $\Phi'$.
\end{enumerate}
The restriction of the polarizations to polarization classes in (a) 
has been discussed in 
Section~\ref{sec:polarizations}. Moreover, choice (b) can be given a quite intuitive picture
coming from Dirac's original idea that the motion deep down in the sea should be
irrelevant when studying the excitations on its ``surface''.  Clearly, when a sea
wave function  $\Wedge \Psi\in\cF_{\Wedge \Phi}$, which could represent an
excitation w.r.t.\  $\Wedge \Phi$, is evolved from $\Sigma$ to $\Wedge \Psi'$ on
$\Sigma'$, clearly also the particles deep down in the sea will ``move''.  Since
there are infinitely many it will be impossible to directly compare $\Psi'$ with
$\Psi$ in general.  Writing $U=U_{\Sigma'\Sigma}^A$ in matrix notation
\begin{align}
    U
    =
    \begin{pmatrix}
        U_{++} & U_{+-} \\
        U_{-+} & U_{--}
    \end{pmatrix}
    =
    \begin{pmatrix}
        P^{{V'}^\perp}_{\Sigma'} U P^{V^\perp}_{\Sigma} 
        & P^{{V'}^\perp}_{\Sigma'} U P^V_{\Sigma} \\
        P^{{V'}}_{\Sigma'} U P^{V^\perp}_{\Sigma}
        & P^{{V'}}_{\Sigma'} U P^{V}_{\Sigma}
    \end{pmatrix},
\end{align}
the motion deep down in the sea is governed by $U_{--}$. Now, if according to
Dirac's original idea the motion deep down in the sea can be considered
irrelevant for the behavior of the excitations on its surface
one should still be able to compare $\Wedge\Psi'$ to $\Wedge \Psi$ when
reversing the motion deep down in the sea with $(U_{--})^{-1}$.  If $U$ is for
example
sufficiently close to the identity this can be done explicitly since then
$U_{--}$ has an inverse $ R=(U_{--})^{-1}$. As we shall see now, the
inversion of the motion deep down in the sea can be implemented by means of an
operation from the right $\cR_{ R}$. For $ R$ to induce an operation
from the right it has to be asymptotically orthonormal, i.e., $ R^* R$
must have a determinant.  Recall that condition (a) in
Theorem~\ref{thm:our-shale} states that the off-diagonals $U_{+-}$ and $U_{-+}$
are Hilbert-Schmidt operators. 
Thanks to $U^*U=\id_{\cH}$ the
identity
\begin{align} 
    U_{--}^*U_{--} = \id_{V} - (U^*)_{-+}U_{+-}
\end{align} 
holds, and
since the product of two Hilbert-Schmidt operators has a trace, one finds
$U_{--}^*U_{--}\in \id_V+I_1(V)$. Thus, $U_{--}^*U_{--}$ and 
then also $ R^* R$ have determinants. 
Note that
in general $\det (R^*R)\neq 1$, which implies that $\cR_R$ may fail to be unitary
up to the factor $\det |R|$.
By definition one finds
\begin{align} 
    \cR_{ R} \circ \cL_U \cF_{\Wedge \Phi}
    =
    \cF_{\Wedge U\Phi  R}=\cF_{\Wedge \Phi'}
\end{align} 
because ${\Phi'}^* U\Phi  R = P^{V'} ( U_{+-} + U_{--})  R
= \id_{ V'}$, and therefore, has a determinant. In consequence, we
yield the unitary Dirac evolution
\begin{align}
    \widetilde U^A_{V,\Sigma;V'\Sigma'}:
    \cF_{\Wedge\Phi}
    \to
    \cF_{\Wedge\Phi'},
    \quad
    \widetilde U^A_{V,\Sigma;V'\Sigma'}
    =
    \det |(U^A_{\Sigma'\Sigma})_{--}|
    \;
    \cR_{[(U^A_{\Sigma'\Sigma})_{--}]^{-1}} \circ \cL_{U^A_{\Sigma'\Sigma}},
\end{align}
which implements both the forward evolution of the whole Dirac sea and the
backward evolution of the states deep down in the sea.

\section{The charge current and the phase of the evolution operator}
\label{sec:phase}

Although the construction of the second-quantized evolution operator according
to the above program is successful, it fails to identify the phase. This
short-coming has no effect on the uniqueness of transition probabilities but
it turns out that the charge current depends directly on this phase.
One way to see that is from Bogolyubov's formula of the current
\begin{align}
    \label{eq:current}
    J^\mu(x)=i \, \widetilde U_{V_\inn,\Sigma_\inn;V_\out\Sigma_\out}^A
    \,
    \frac{\delta}{\delta A_\mu(x)}
    \,
    \widetilde U_{V_\out,\Sigma_\out;V_\inn,\Sigma_\inn}^A,
\end{align}
where $\Sigma_\out$ is a Cauchy surface in the remote future of the support of
$A$ such that $\Sigma_\out\cap\supp A=\emptyset$. Changing the evolution
operator by an $A$-dependent phase generates another summand on the right hand
side of \eqref{eq:current} by the chain rule. Until some phase is distinguished,
\eqref{eq:current} has no particular physical meaning as charge current.
\red{Nevertheless, all possible currents can be derived from \eqref{eq:current}
given an evolution operator and a particular phase.} Therefore, the situation is
better than in standard QED. There, the charge current is a quantity whose formal
perturbation series leads to several divergent integrals which have to be taken
out by hand until only a logarithmic divergent is left, which in turn is
remedied by means of charge renormalization.  On the contrary, here, the
currents are well-defined and in a sense the correct one only needs to be
identified by determining the phase of the evolution operator.  As already
envisioned in \cite{scharf_finite_1995} and discussed by
\cite{Mickelsson1998,gracia2000}, this may be done by imposing extra
conditions on the evolution operator. One of them is clearly the following
property.  For any choice of a future oriented foliation of space-time into a
family of Cauchy surfaces $(\Sigma_t)_{t\in\R}$ and polarization $V_t\in
C_{\Sigma_t}(A)$, $t\in\R$, the assigned phase of the evolution operator
$\widetilde U^A(t_1,t_0)=\widetilde
U^A_{\Sigma_{t_1},V_{t_1};\Sigma_{t_0},V_{t_0}}$ constructed in
Section~\ref{sec:wedge-spaces} should be required to fulfill $\widetilde
U(t_1,t_0)=\widetilde U(t_1,t)\widetilde U(t,t_0)$. Other constraints come from
the fact that $J^\mu(x)$ must be Lorentz and gauge covariant, and its vacuum
expectation value for $A=0$ should be zero. The hope is that the collection of
all such physical constraints restrict the possible currents \eqref{eq:current} to a
class which can be parametrized by a real number only, the electric charge of
the electron.  In the case of equal-time hyperplanes one possible choice of the
phase was given by Mickelsson via a parallel transport argument
\cite{mickelsson_phase_2014}. On top of the nice geometric construction and
despite the fact that there are still degrees of freedom left, Mickelsson's
current agrees with conventional perturbation theory up to second order. The aim
of this program is to settle the question which conditions are required to
identify the charge current upon changes of the value of the electric charge.\\

\thanks{\textbf{Acknowledgment:} This work has partially been funded by the Elite
Network of Bavaria through the JRG ``Interaction between Light and Matter''.}



\begin{thebibliography}{10}

\bibitem{Anderson1933}
C.D. Anderson,
\textit{The Positive Electron.}
{The Physical Review} \textbf{43}:6 (1933), 491--494.

\bibitem{baer:2007}
C.~Bär, N.~Ginoux, and F.~Pfäffle,
\textit{Wave Equations on Lorentzian Manifolds and Quantization}.
European Mathematical Society, 2007.

\bibitem{Deckert2010}
D.-A. Deckert,
\textit{{Electrodynamic Absorber Theory -- A Mathematical Study}},
{Der Andere Verlag}, 2010.

\bibitem{Deckert2010a}
D.-A. Deckert, D.~D\"{u}rr, F.~Merkl, and M.~Schottenloher,
\textit{Time-evolution of the external field problem in Quantum
  Electrodynamics.}
{{Journal of Mathematical Physics}} \textbf{51}:12 (2010), 122301.

\bibitem{Deckert2014}
D.-A. Deckert and F.~Merkl,
\textit{Dirac equation with external potential and initial data on Cauchy
  surfaces.}
{Journal of Mathematical Physics} \textbf{55}:12 (2014), 122305.

\bibitem{Deckert2015}
D.-A. Deckert and F.~Merkl,
\textit{External field QED on Cauchy surfaces.}
In preparation.

\bibitem{Derezinski:13}
J.~Dereziński and C.~Gérard,
\textit{Mathematics of Quantization and Quantum Fields.}
Cambridge University Press, 2013.

\bibitem{dimock:82}
J.~Dimock,
\textit{Dirac quantum fields on a manifold}.
{{Transactions of the American Mathematical Society}}
  \textbf{269}:1 (1982), 133--147.

\bibitem{Dirac1934}
P.A.M. Dirac,
\textit{Theorie du Positron.}
{Selected Papers on Quantum Electrodynamics Edited by J.
  Schwinger, Dover Publications Inc., New York}, 1934.

\bibitem{Fierz1979}
H.~Fierz and G.~Scharf,
\textit{Particle interpretation for external field problems in QED.}
{Helvetica Physica Acta. Physica Theoretica},
  \textbf{52}:4 (1980), 437--453.

\bibitem{FinsterBook}
F.~Finster,
\textit{The {C}ontinuum {L}imit of {C}ausal {F}ermion {S}ystems.}
Book based on the preprints arXiv:0908.1542,
  arXiv:1211.3351, and arXiv:1409.2568, in preparation.

\bibitem{Finster2015a}
F.~Finster and J.~Kleiner,
\textit{Causal {Fermion} {Systems} as a {Candidate} for a {Unified}
{Physical} {Theory}.}
arXiv:1502.03587.

\bibitem{Finster2015b}
F.~Finster, J.~Kleiner, and J.-H. Treude,
\textit{An {I}ntroduction to the {F}ermionic {P}rojector and {C}ausal
{F}ermion {S}ystems.}
{In preparation}.

\bibitem{Finster:2012}
F.~Finster, J.~Kleinert, and J.H. Treude,
\textit{An introduction to the fermionic projector and causal fermion
systems.}
Transactions of the American Mathematical Society (2012).

\bibitem{gracia2000}
J.M. Gracia-Bondia,
\textit{The phase of the scattering matrix.}
{Physics Letters B} \textbf{482}:1-3 (2000), 315--322.

\bibitem{Gravejat2013}
P.~Gravejat, C.~Hainzl, M.~Lewin, and E.~S\'{e}r\'{e},
\textit{Construction of the Pauli\textendash Villars-Regulated Dirac Vacuum
  in Electromagnetic Fields.}
{Archive for Rational Mechanics and Analysis}
  \textbf{208}:2 (2013), 603--665.

\bibitem{greiner_relativistic_2000}
W.~Greiner and D.A. Bromley,
\textit{Relativistic Quantum Mechanics. Wave Equations.}
Springer, Berlin; New York, 3rd edition, 2000.

\bibitem{john:82}
F.~John,
\textit{Partial Differential Equations.}
Springer, New York, 1982.

\bibitem{kay_theorems_1991}
B.~S. Kay and R.~M. Wald,
\textit{Theorems on the uniqueness and thermal properties of stationary,
nonsingular, quasifree states on spacetimes with a bifurcate killing horizon.}
{Physics Reports} \textbf{207}:2 (1991) 49--136.

\bibitem{Klein:1929}
O.~Klein,
\textit{Die Reflexion von Elektronen an einem Potentialsprung nach der
  relativistischen Dynamik von Dirac.}
  {Zeitschrift für Physik} \textbf{53}:3-4 (1929), 157--165.

\bibitem{Langmann1996}
E.~Langmann and J.~Mickelsson,
\textit{Scattering matrix in external field problems.}
{{Journal of Mathematical Physics}} \textbf{37}:8 (1996), 3933--3953.

\bibitem{Mickelsson1998}
J.~Mickelsson,
\textit{Vacuum polarization and the geometric phase: gauge invariance.}
{{Journal of Mathematical Physics}} \textbf{39}:2
  (1998), 831---837.

\bibitem{mickelsson_phase_2014}
J.~Mickelsson,
\textit{The phase of the scattering operator from the geometry of certain
infinite-dimensional groups.}
      {Letters in Mathematical Physics} \textbf{104}:10 (2014), 1189--1199.

\bibitem{Pickl2008}
P.~Pickl and D.~D\"{u}rr,
\textit{Adiabatic pair creation in heavy-ion and laser fields.}
{Europhysics Letters} \textbf{81}:4 (2008), 40001.

\bibitem{ringstroem:09}
H.~Ringström,
\textit{The Cauchy problem in general relativity.}
European Mathematical Society, Zürich, 2009.

\bibitem{Ruijsenaars1977a}
S.N.M. Ruijsenaars,
\textit{Charged particles in external fields. I. Classical theory.}
{{Journal of Mathematical Physics}} \textbf{18}:4
  (1977), 720--737.

\bibitem{Ruijsenaars1977}
S.N.M. Ruijsenaars,
\textit{Charged particles in external fields. II. The quantized Dirac and
  Klein-Gordon theories.}
{{Communications in Mathematical Physics}}
  \textbf{52}:3 (1977), 267--294.

\bibitem{scharf_finite_1995}
G.~Scharf,
\textit{Finite Quantum Electrodynamics: The Causal Approach.}
Springer, Berlin; New York, 2nd edition, 1995.

\bibitem{Schroedinger:1930}
E.~Schr{\"o}dinger,
\textit{{\"U}ber die kr{\"a}ftefreie Bewegung in der relativistischen
  Quantenmechanik.}
{Berliner Ber.} (1930), 418–428.

\bibitem{Shale1965}
D.~Shale and W.~F. Stinespring,
\textit{Spinor representations of infinite orthogonal groups.}
{{Journal of Mathematics and Mechanics}} \textbf{14} (1965), 315--322.

\bibitem{taylor:11}
M.E. Taylor,
\textit{Partial differential equations {III}.}
Springer, New York, 2011.

\end{thebibliography}
\end{document}